\documentclass[aps,prl,twocolumn]{revtex4-1}
\usepackage{amsmath}
\usepackage{amssymb}
\usepackage{graphics}
\usepackage{epsfig}
\usepackage{color}
\usepackage{graphicx}
\usepackage{color}
\begin{document}
\title{Ecosystems with mutually exclusive interactions self-organize to a state of high diversity}
\author{Joachim Mathiesen, Namiko Mitarai, Kim Sneppen and Ala Trusina}

\affiliation{Niels Bohr Institute / CMOL,  Blegdamsvej 17, DK-2100 Copenhagen, Denmark}

\begin{abstract}
Ecological systems comprise an astonishing diversity of species that cooperate or
compete with each other forming complex mutual dependencies.
The minimum requirements to maintain a large species diversity 
on long time scales are in general unknown. 
Using lichen communities as an example, we propose 
a model for the evolution of mutually excluding organisms that compete for space.
We suggest that chain-like or cyclic invasions involving
three or more species open for creation of spatially separated sub-populations 
that subsequently can lead to increased diversity.
In contrast to its non-spatial counterpart,
our model predicts robust co-existence of a large number of species,
in accordance with observations on lichen growth.
It is demonstrated that large species diversity can be obtained on
evolutionary timescales, provided that interactions between species have spatial constraints.
In particular, a phase transition to a sustainable state of high diversity
is identified.
\end{abstract}

\maketitle

{\it Introduction.}--
Interactions between biological species may well be as old as life itself
\cite{eigen,tejedor,smith} with competition and predation 
 as major determinants 
for species diversity \cite{dayton,paine}.
Competitive exclusion \cite{gause,hardin} has been suggested to reduce ecosystem diversity when several species compete for the same resources. Real ecosystems, on the other hand, consist of multiple species and have a robustness
that may even increase with diversity \cite{tilman,ives}. To obtain robustness of ecosystem diversity in theoretical models one needs first of all to limit the
exponential growth by assuming a maximum carrying capacity for the 
population of each species \cite{vandermeer}. Extreme version of such models \cite{jain,segre}
indeed predicts a sustainable but fragile co-existence of multiple species.

A more robust way to maintain high diversity is to include space  \cite{mollison,hufkens,kareiva,kerr,schrag,heilmann}
e.g.~in combination with hypercycles \cite{bjoerlist} or predator-prey cycles \cite{gilpin,Perc2007, Laird2006,Szabo2001}. 
Studying ecosystems in marine hard-substrate environments Jackson \& Buss \cite{jackson} suggested that non-transitive allelopathic relationships between species could maintain species diversity on a longer timescale than pure hierarchical predation relationships. This was confirmed in a model on two-dimensional lattice where sessile species compete for space\cite{buss,karlson}.
\begin{figure}[htb]
\centerline{\includegraphics[width=.35\textwidth]{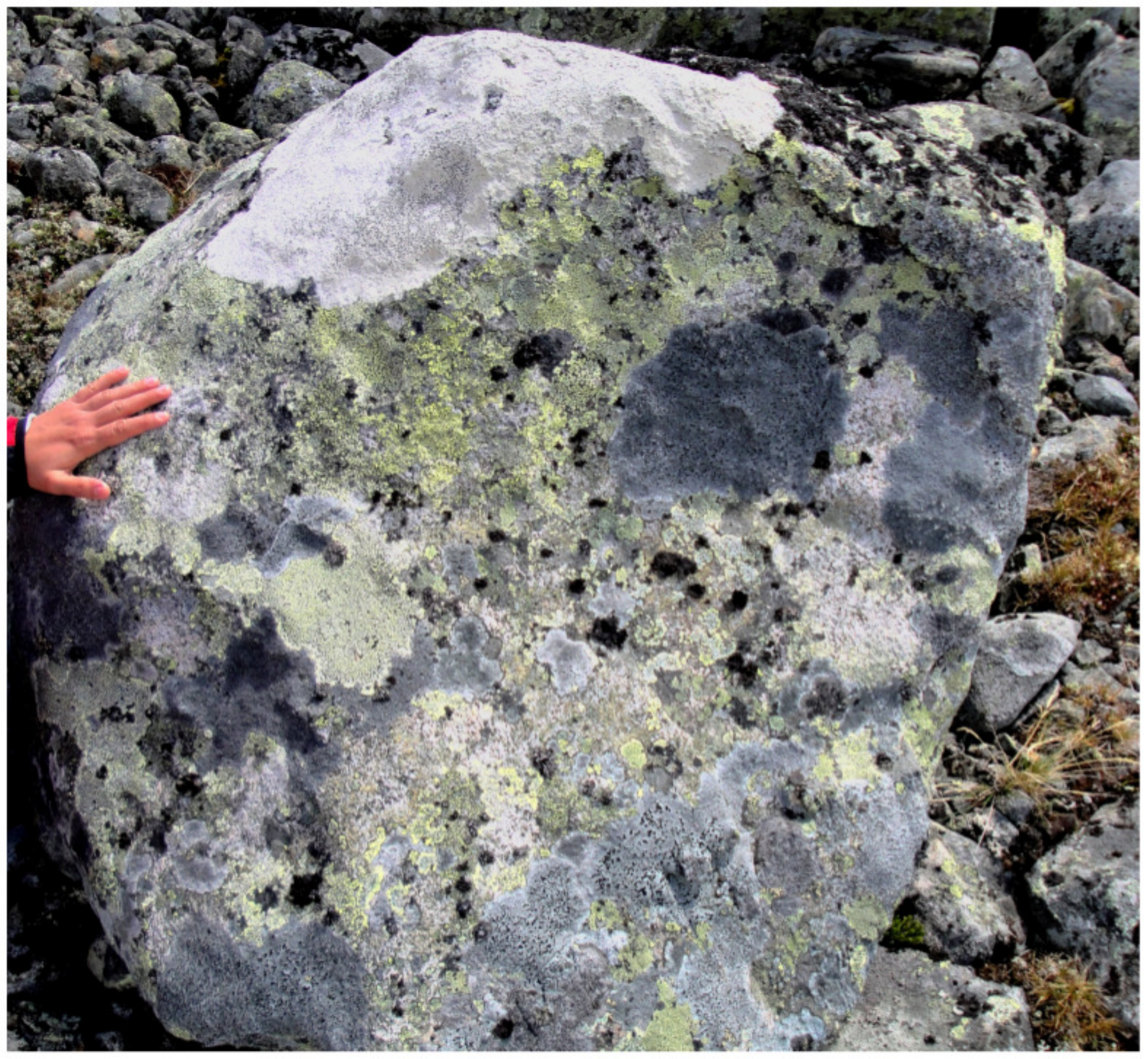}}
\caption{\small\sl Photograph of a crustose lichen community on a rock in an alpine environment (at 1300m altitude, Jotunheimen, Norway).
}\label{figure1}
\end{figure}%
As in the ''Buss`` model \cite{buss,karlson} we consider a 2-dimensional lattice where the competition for resources 
is a zero sum game about available space.
In our model, however, the focus is on the dynamic balance between an introduction of new species
and exclusion of older species.
With this complementary model we, for the first time show that:
a) There is a sharp transition from multiple to single species as the number of
 interactions is increased.
b) Both cycles and chain-like (hierarchical) relationships lead to spatial fragmentation of species population
thus creating isolated niches for new species and increased  diversity.
 
The model is inspired by the spatial dynamics of lichen communities. 
Lichens have existed for as long as $\sim$600 million years \cite{YXT05} and are organisms 
consisting of fungi and algae living in a symbiosis \cite{N08}. 
Communities of lichens are formed by a combination of slow local growth and 
a reproductive strategy where fungal spores or propagules containing 
the intact symbiosis are dispersed (e.g. by winds or water flows) over 
lengthscales much larger than the size of the communities \cite{MFCBM04}. 
Because of this long distance dispersal, we assume that new species come from far away, and are completely
unrelated to any of the species that already colonize our system.
\begin{figure}[ht]
\centerline{\includegraphics[width=.40\textwidth]{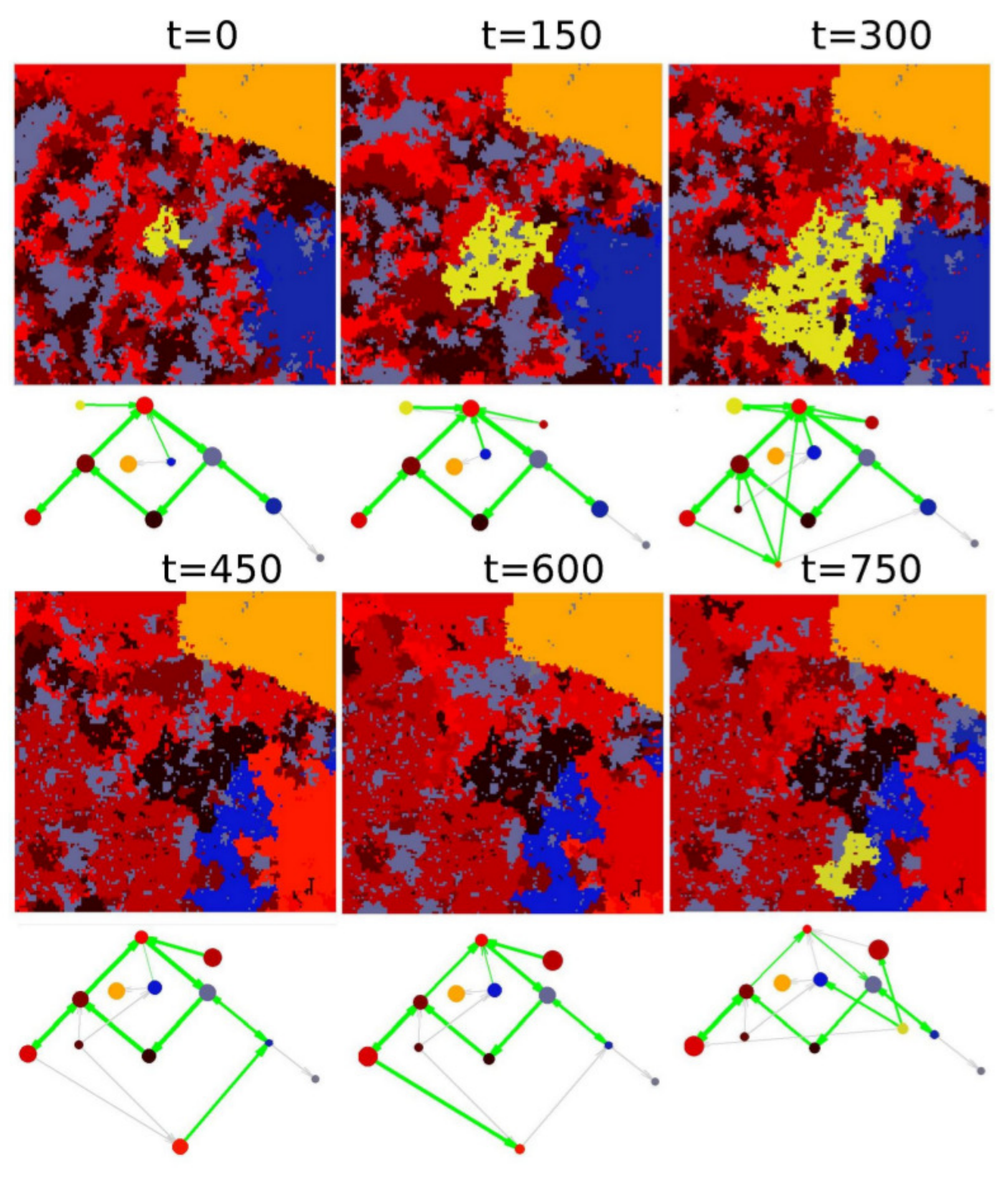}}
\caption{\small\sl Visualization of six successive snapshots of a simulation with $\gamma=0.1$ and $\alpha=0.1$ for a system of size $L=200$.
The networks below the snapshots illustrate the actual (green) and potential (grey) links between the species in the system. The size of a node represents the current population size of the corresponding species whereas the thickness of green links
quantifies the number of active invasions sites.
}\label{figure2}  
\end{figure}

Fig.~\ref{figure1} shows a crustose lichen community in an alpine environment. These lichens 
grow about $0.1$mm/year and typically covers a rock surface that 
has recently been exposed on a timescale  
$\sim 100\rightarrow 1000$ years \cite{JNHTM10}. When a crustose lichen meets another, a contact boundary is formed, 
and if they are competitively equal the boundary remains stable over time. 
The bulging boundaries between various species seen in Fig.~\ref{figure1} suggest a 
mini-ecosystem with complex interactions. The interaction of these species may be 
represented by a directed network with directions from superior to inferior species. 
This network does not necessarily have any particular species as the most fit one 
in agreement with the fact that any sizable rock typically hosts multiple lichen species.

{\it Results.--}
Our model considers a multi-species ecosystem competing on 
a two dimensional lattice of sites which at any given time can be occupied by one species only. 
The species on one site can invade a neighbor site,
provided that it is occupied by a competitively inferior species. 
The emerging ecosystem can be characterized by a directed network 
of possible species interactions. Interactions which are materialized 
only when organisms of the respective species are neighbors somewhere in the system. 
The aim of our model is to study ecosystem diversity as we change 
the number of potential interactions between species, parametrized by $\gamma$.
In addition to this, new species 
are introduced with a rate $\alpha$. 
\begin{figure}[ht]
\centerline{\includegraphics[width=.35\textwidth]{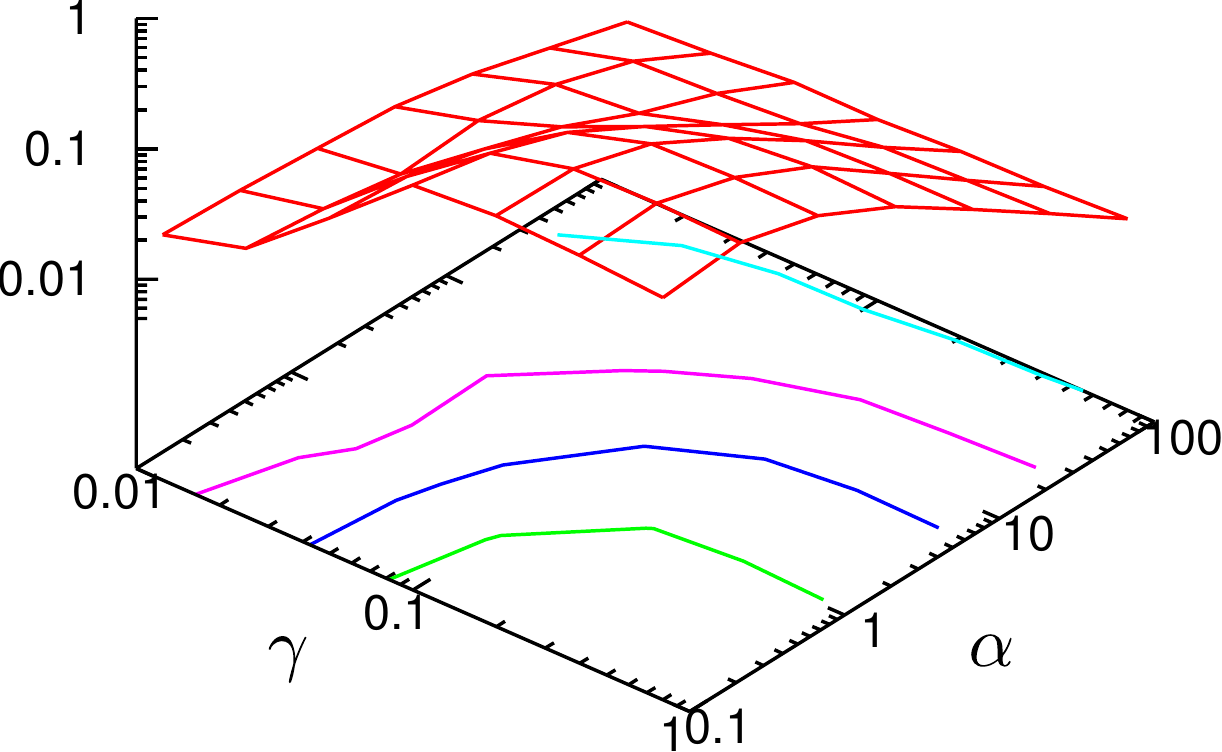}}
\caption{\small\sl Active interactions between species in the system. The number of interactions are measured in units of the maximum potential interactions (sum of all outgoing links in the network) and shown as function of $\gamma$ and $\alpha$.
The computations were carried out on a system of size $L=200$.
}\label{network}
\end{figure}
Each time step of our model consists of two possible events:
(i)
Select a random site $i$ and one of its nearest
neighbors $j$. If the species $s(i)$ at site $i$ can invade the species
$s(j)$ at site $j$, i.e. $\Gamma (s(i),s(j))=1$, then site $j$ is
updated by setting $s(j)=s(i)$. Here  $\Gamma$ is the matrix that 
represents the possible interactions. These interactions remain fixed once they are introduced. 
(i)
With probability $\alpha \times \gamma /N$ a new random species $s$ is introduced
at a random point $j$ and assigned random interactions $\Gamma (s,u)$ 
and $\Gamma (u,s)$ with
all existing species $u$ in the system. Each of these interactions are 
assigned value 1 with probability $\gamma$, or otherwise set to $0$ (we do allow for the case $\Gamma (u,s)=\Gamma (s,u)=1$).
The introduced species $s$ is assumed to be able to invade the previous species
at the site $j$, $s(j)$: $\Gamma (s,s(j))=1$.

In Fig.~\ref{figure2}, we show snapshots of a
model ecosystem of size $N=L\times L=200\times 200$ using open boundary conditions.
The snapshots represent a typical steady state behavior using $\gamma=0.1$ and $\alpha=0.1$.
The different colors represent different species. 
Parts of the lattice are inactive (e.g. the orange colored species in the upper right corner), while other parts are exposed to active cycles, like the 4 mutually invading species
$grey \rightarrow$   $dark$ $brown \rightarrow$   $brown \rightarrow$   $red \rightarrow grey$.
This cycle is also illustrated in terms of a network in the lower panels where the thickness of green links represents the number of active invasion events per time unit.

With time the diversity (number of species in the system) $D$ reaches a steady state where there is a balance between new species eliminating more than one of its prey species, and an increase in $D$
when a new species only invades a fraction of an existing species. An increase in diversity requires that the population of existing species are fragmented into spatially separate regions. 
{Two locally interacting species cannot increase spatial heterogeneity: 
One species will eventually be replaced by the other. 
To generate spatially separated regions,
one needs simultaneous dynamics of at least 3 species locally.
Such multiple interactions may be both chain-like or cyclic. 
For example a species A that emerges inside the territory of B,
can result in a fragmentation of B if a third species C can invade A but not B. 
Whether B invades C or not, distinguishes a cycle from a chain-like
relationship, but both can generate spatial fragmentation.
However the cyclic relationship tends to live longer.}


In the network $\Gamma$ a link is called active if it represents an interaction between two species which are physically in contact. For low $\gamma$ only a small fraction of the links $\Gamma (s,u)=1$ are active since the species $s$ and $u$ are often physically separated by other species which neither can invade. For larger $\gamma$ we observe fewer and more widespread species which are often in contact leading to a larger fraction of active links, see Fig.~\ref{network}. 

\begin{figure}[ht]
\centerline{\epsfig{file=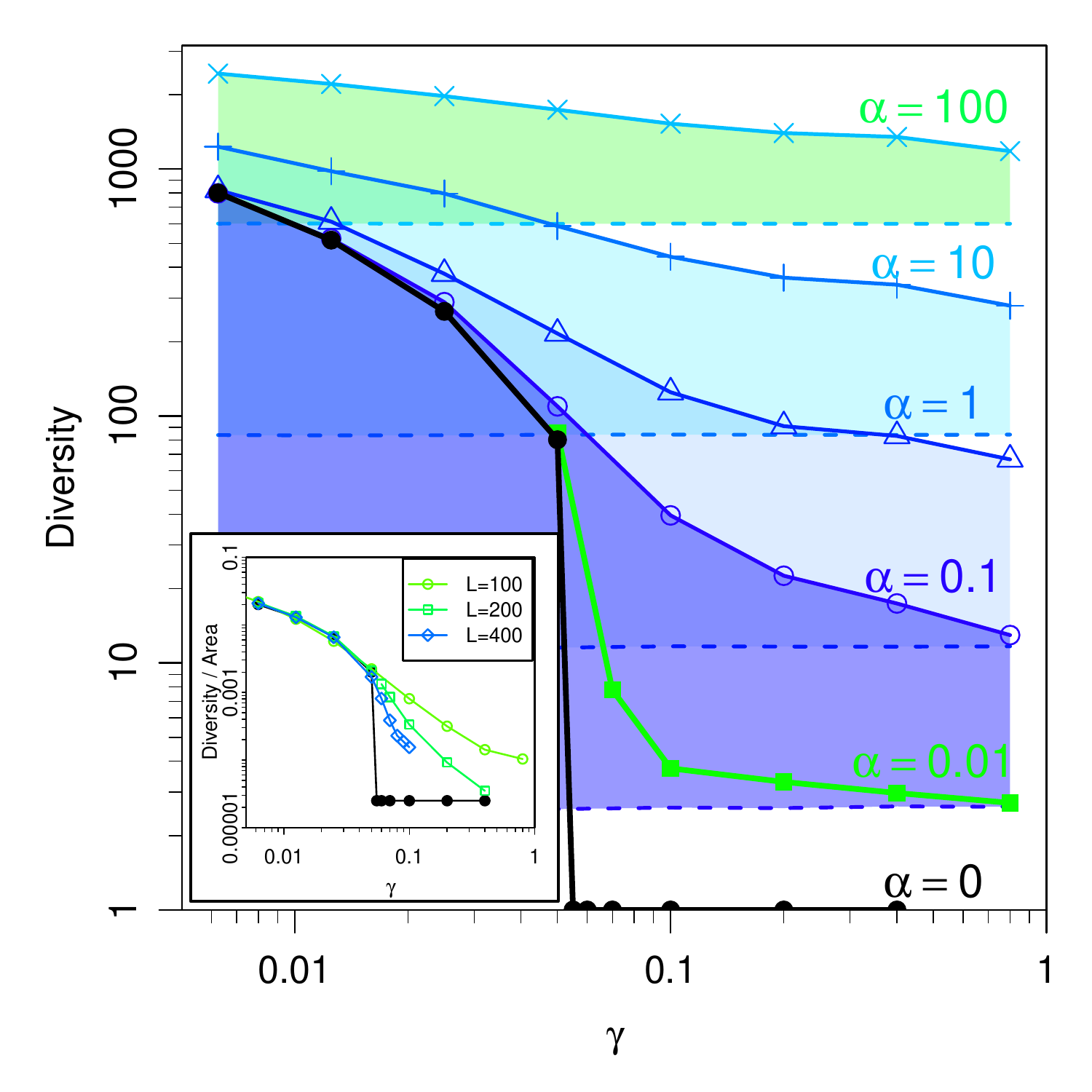,width=.48\textwidth}}
\caption{\small\sl Diversity $D$ as function
of $\gamma$ for different $\alpha$ values. The shaded areas mark the difference between the spatial model (upper bounding lines) and the random neighbor version (dashed lower lines). For the non-spatial version the diversity
$D \approx 11\times \alpha^{0.85}+1$ for $L=200$
is independent of $\gamma$. The black line shows the diversity achieved from the quasistatic limit. The inset shows simulations in the quasistatic limit where two species are simultaneously introduced whenever a configuration freezes. The black line is identical to the quasistatic limit shown in the main panel. The inset also shows that $D$/Area acts as an order parameter for the system.
}\label{divcomp}
\end{figure}

The central feature of the model is the ability to 
sustain high diversity, even in the limit $\alpha \rightarrow 0$.
Fig.~\ref{divcomp} quantifies this for a $N=200 \times 200$ system,
which shows the finite diversity $D$ in the limit of $\alpha\to 0$
for $\gamma< 0.055$. In this figure we also show $D(\gamma)$
obtained from a quasistatic simulation where new species are only introduced 
when there is no ongoing population dynamics. 
For $\gamma>\gamma_c \approx 0.055$ the quasistatic diversity is $D(\gamma)=1$.

The quasistatic version of the model
reflects the biologically interesting limit where evolution rates are 
much slower than any dynamics associated to the species populations. 
The quasistatic simulation has two 
metastable states when $\gamma<\gamma_c$, one with a high but 
finite diversity and one absorbing state where $D=1$. 
In the simulations, the state of high diversity is reached by starting 
from a state with large diversity. In simulating the quasistatic dynamics
there will occasionally be long periods where several species compete
dynamically for the same area. To shorten these periods we eliminate all outgoing links from one randomly chosen active species. If this does not stop the dynamics, the elimination procedure is repeated until the system eventually freezes. 
After the system has frozen, the eliminated links are re-introduced and the quasistatic
simulation is continued by introducing a new random species.
The obtained steady state diversity (black line in Fig.~\ref{divcomp}) is close
to the full simulation at $\alpha=0.01$. 

Importantly, the obtained large diversity 
at $\gamma < \gamma_c$ 
is also found in an extended
model where species sometimes compete for free space created by occasional
death of individuals, i.e. the system behaviour 
is robust to external pertubations of the type considered in \cite{karlson}. 
In the quasistatic limit the sharp transition to high diversity is maintained 
even when 10\% of the systems sites 
are killed prior to the introduction of each new species. For finite $\alpha$,
occasional death will soften the transition.  

In the insert of Fig.~\ref{divcomp}, we examine a variant of the quasistatic model
where two new species are introduced simultaneously instead of only one.
Thereby the absorbing state at $D=1$ disappears and the previously
metastable state at high diversity becomes a true steady state.
One sees that the obtained steady state diversity for this variant of the model 
is close to that obtained from the $\alpha\rightarrow 0$ limit of the standard model.
Also, the insert in Fig.~\ref{divcomp} shows that the 
behavior of the one- and two-species introduction models have a similar 
high diversity state when $N\rightarrow \infty$. 
Both models exhibit a phase transition to the high diversity state for $\gamma<\gamma_c\approx 0.055$.
The transition is maintained also when we allow the invasion rates 
to vary between the species, as well as when we implement the model
on a triangular lattice.
Close to the critical point, the finite but small $\alpha$ simulations
exhibits pronounced bi-stability with rare transitions between a high and low
diversity state. Transitions from low to high diversity are created through a state with many disconnected patches of a few species, whereas the breakdown of high diversity occurs when each species is only represented by a few patches. 

For real sessile ecosystems, $\gamma$ can be estimated from fraction of active boundaries, which have been reported down to a value of $2.5\%$ for sponge-corals competition \cite{AS1997}. The low $\gamma$ is also consistent with observations on crustose lichens \cite{JNHTM10}.

{\it Discussion.--}
The model results in an interesting interplay between an interaction network \cite{rosvall,sole} and spatial dynamics. That is, the spatial configuration of the species limits the number of active links in the network and as a results most links are passive and have no influence on the dynamics. This is in contrast to the common network assumption that all links represent ongoing interactions. In biological systems most links are transient. To explore further the influence of the space-network coupling, we have compared our spatial model with an equivalent random-neighbor version where lattice sites interact at random and independent of their spatial locations. The non-spatial model is a noisy version of average population equations of the form
$\frac{dX_i}{dt} = \sum_j \Gamma(i,j) X_i X_j  - \sum_j \Gamma(j,i) X_i X_j,$
with the additional rule that species with near zero population are removed. By comparing the dashed and solid lines in Fig.~\ref{divcomp}, we see that the behavior of the random-neighbor model differs
from our model,
(i) $D$ is independent of $\gamma$ and  in general much lower than for the spatial model.
In the quasistatic limit then $D=1$ for all $\gamma$ values.
(ii)
At moderate $\alpha$ and $\gamma$ the random neighbor model
can develop sustained oscillatory states, resembling the hypercycles 
of Eigen and Shuster \cite{eigen}. The size and duration of these cycles
decrease with the rate of introduction of new species $\alpha$.
In the spatial model these cycles emerge in restricted spatial regions,
cause redistribution of species boundaries and give rise to sub-populations separated by neutral species.

The process that increases diversity in the model speaks to an 
interesting interplay between a sympatric and an allopatric speciation process. 
Although we have not implemented genealogical species relationships, 
the algorithm might be seen as a minimal evolution model, 
where new species are related to the species in the region where they originate. 
In this framework, the spatial  division of an existing species population by a 
neutral species and the subsequent invasion of one fragment by a new species 
then amount to an allopatric speciation event. 
Overall, our model can create two species in a region previously covered by one species.

The main feature of our spatial model is the emergence of a robust ecological system with multiple co-existing species. 
The species diversity depends crucially on both the limitations imposed by the interaction network, 
and on the spatial positioning of the individual species. If either space or network constraint 
is absent the ecosystem complexity cannot be maintained. 
The co-existence is dynamic in the sense that it includes both extinction as well as occasional re-emergence of new species.
In \cite{buss} it was found that non-transitive relationships 
indeed prolonged co-existence of many species. However in that model a
substantial fraction of the lattice was often cleared by external disturbances, 
and any of the initially present 10 species were repeatedly introduced on empty lattice positions.
In contrast by allowing introduction of species with new random interactions,
our model demonstrated a self-organization towards an 
ecosystem with substantial diversity 
provided that the interaction probability $\gamma<\gamma_c\sim 0.055$.
{ A system diversity which in fact remains even when we artificially cut cyclic relationships
in the quasistatic limit, demonstrating that steady state species diversity crucially depends on chain-like spatial 
interactions that involves at least 3 species.}

In an evolutionary perspective, an interesting aspect of our
model is the fact that it does not rely on any ultimate fitness landscape.
At any point in space and time, the "fitness" of a species is dictated by 
its neighborhood \cite{kauffman,baksneppen}. 
Our model supplements these studies by proposing
a self-organized allopatric speciation mechanism which suggest a minimal sustainable diversity  
for ecosystems of mutually exclusive species in 2-dimensions.

{\it Aknowledgement} This study was supported by the Danish National Research Foundation through the Center for Models of Life.
\bibliographystyle{h-physrev3}
\bibliography{plants}

\end{document}